\documentclass[runningheads]{llncs}
\usepackage{booktabs}
\usepackage{multirow}
\usepackage{amsfonts,amssymb}
\usepackage{hyperref}
\usepackage{cite}
\usepackage{float}
\hypersetup{
    colorlinks=true,
    anchorcolor=red,
    linkcolor=red,
    urlcolor=blue,
    citecolor=green,
}\usepackage{caption}
\usepackage{graphicx}

\begin{document}
\title{CU-Net: Cascaded U-Net with Loss Weighted Sampling for Brain Tumor Segmentation}

\titlerunning{Cascaded U-Net with LWS for Brain Tumor Segmentation}
%
\author{Hongying Liu\inst{1} \and Xiongjie Shen\inst{1} \and Fanhua Shang\inst{1} \and
Fei Wang\inst{2}}
\authorrunning{H. Liu, X. Shen, F. Shang, F. Wang.}
\institute{Key Lab of Intelligent Perception and Image Understanding of Ministry
of Education, School of Artificial Intelligence, Xidian University, China \\ \email{\{hyliu, fhshang\}@xidian.edu.cn, shenxiongjie123@gmail.com}
\and Weill Cornell Medical School, Cornell University, NY, USA
\email{feiwang.cornell@gmail.com}}

\maketitle

\begin{abstract}
This paper proposes a novel cascaded U-Net for brain tumor segmentation. Inspired by the distinct hierarchical structure of brain tumor, we design a cascaded deep network framework, in which the whole tumor is segmented firstly and then the tumor internal substructures are further segmented. Considering that the increase of the network depth brought by cascade structures leads to a loss of accurate localization information in deeper layers, we construct many skip connections to link features at the same resolution and transmit detailed information from shallow layers to the deeper layers. Then we present a loss weighted sampling (LWS) scheme to eliminate the issue of imbalanced data during training the network. Experimental results on BraTS 2017 data show that our architecture framework outperforms the state-of-the-art segmentation algorithms, especially in terms of segmentation sensitivity.
\keywords{Brain tumor segmentation  \and Cascaded U-Net \and  Feature fusion  \and  Loss weighted sampling.}
\end{abstract}

\section{Introduction}
Glioma is the most common primary central nervous system tumor with high morbidity and mortality. For glioma diagnosis, four standard Magnetic Resonance Imaging (MRI) modalities are generally used: T1-weighted MRI (T1), T2-weighted MRI (T2), T1-weighted MRI with gadolinium contrast enhancement (T1ce) and Fluid Attenuated Inversion Recovery (FLAIR). In fact, it is challenging and time-consuming for doctors to combine these four modalities to complete a fine segmentation of brain tumors.

Since deep learning has attracted considerable attentions from researchers, convolutional neural network (CNN) has been widely applied to the brain tumor segmentation. Havaei et al.~\cite{havaei2017brain} proposed a CNN architecture with two pathways to extract features in different scales. Such an idea of multi-scale was validated to be effective in improving the segmentation results in many works~\cite{havaei2017brain,kayalibay2017cnn,havaei2015convolutional}. In \cite{wang2017automatic}, a triple cascaded framework was put forward according to the hierarchy of brain tumor, though novel in framework, the patch-wise and sequential training process leads to a somewhat inefficient processing. Shen et al.\ \cite{shen2017multi} built a tree-structured, multi-task fully convolutional network (FCN) to implicitly encode the hierarchical relationship of tumor substructures. The end-to-end network structure was much efficient than the patch-based methods. To improve the segmentation accuracy of tumor boundaries, Shen et al.\ \cite{shen2017boundary} proposed a boundary-aware fully convolutional network (BFCN) and constructed two branches to learn two tasks separately, one for tumor tissue classification and the other for tumor boundary classification. However, the flaw inherent in the traditional FCN still exists, that is, a series of up-samplings after convolution can not attain refined segmentation. To avoid the loss of location information caused by down-sampling operations in traditional CNNs, Lopez et al.~\cite{lopez2017dilated} designed a dilated residual network (DRN) and abandoned pooling operations. This may be a considerable solution to prevent the network from losing the details, but is too time-consuming and memory-consuming.

Ronneberger et al.\ \cite{ronneberger2015u} proposed a U-shape convolutional network (called U-Net) and introduced skip-connections to fuse multi-level features, so as to help the net decode more precisely. Many experimental results show that U-Net performs well in various medical image segmentation tasks. Dong et al.\ \cite{dong2017automatic} applied U-Net to brain tumor segmentation and took the soft dice loss as loss function to solve the issue of imbalanced data in brain MRI data. Though soft dice loss may have better performance than cross entropy loss in some extremely class-imbalanced situation, it has less stable gradient, which may make the training process unstable even not convergent.

Inspired by the hierarchical structure within the brain tumor, we propose a novel cascaded U-shape convolutional network to realize a multistage segmentation of brain tumors. To mitigate the gradient vanishing problem caused by the increase of network depth, each basic block is designed as a residual block. Moreover, we add the decoding-layer supervision during training process and further alleviate the problem of gradient vanishing. To reduce the information loss in the deeper layers, we design many skip connections to help the transmission of high resolution information from the shallow layers to the corresponding deeper layers, so as to obtain more refined segmentation results. To solve the class-imbalanced problem, we present a loss weighted sampling scheme.

The main contribution of this paper can be summarized as follows.
\vspace{-2mm}
\begin{enumerate}
  \item We propose a novel cascaded U-Net with between-net connections for brain tumor segmentation. In particular, each basic block of our cascaded U-Net is designed as a residual block.
  \item We also design many skip connections to help the transmission of high resolution information from shallow layers to deeper layers.
  \item Moreover, we present a loss weighted sampling scheme to address the severe class imbalance problem. The full implementation and the trained networks are available at the authors' website.
  \item Finally, our experimental results show that our method performs much better than state-of-the-art methods in terms of dice score and sensitivity.
\end{enumerate}

\section{The Proposed Cascaded U-Net Method}
\subsection{Our Cascaded U-Net Architecture}
\begin{figure*}[h]
\setlength{\belowcaptionskip}{-0.1cm}
\centering
\includegraphics[width=0.99\columnwidth]{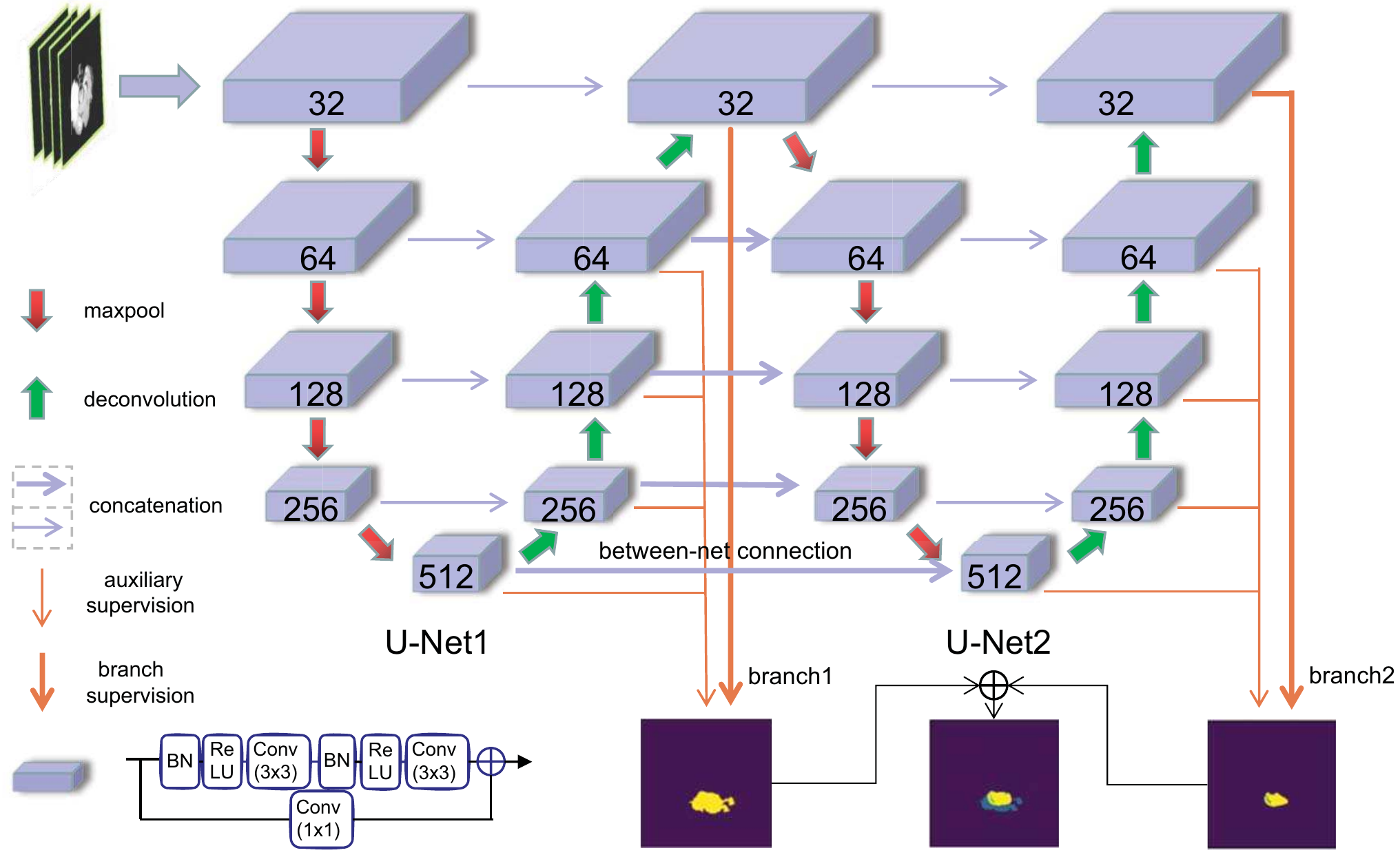}
\vspace{2mm}
\caption{Our cascaded U-Net architecture (CU-Net or CUN) for brain tumor segmentation. The digital number on each block denotes the number of output channels. Before every supervision, including 8 auxiliary supervisions and 2 branch supervisions, there is a $1\times1$ convolution to squeeze the channels of output into the same quantity as target. Besides, in each auxiliary supervision, a deconvolution is used to to up-sample the feature maps to the same resolution as input. All the arrows denote the different operations.}
\label{fig1}
\end{figure*}

Our network is a novel end-to-end architecture. In particular, our cascaded U-Net is mainly composed of two cascaded U-Nets, which are for different tasks, as shown in Fig.~\ref{fig1}. Such a cascaded framework is inspired by the underlying hierarchical structure within the brain tumor. The tumor comprises a tumor core, and the tumor core contains an enhancing tumor.

Given the input brain MRI images, we extract a non-brain mask firstly and prevent the network from learning the masked areas by loss weight sampling. Then the first-stage U-Net separates the whole tumor from background, and sends the extracted features into the second-stage U-Net, which further segments tumor substructures. Such a cascade structure is designed to take advantage of the underlying physiological structure within the brain tumor. The cascade structure will multiply the network depth, which on the one hand will enhance the ability of a network to extract semantic features, but on the other hand exacerbate the gradient vanishing problem. In our architecture, we design the following three strategies to avoid the above problem and fulfill the coarse-to-fine segmentation of brain tumor.

Firstly, inspired by the residual network, each basic unit in our network is constructed by a residual block stacked by two $3\times3$ convolution blocks. Secondly, the auxiliary supervisions are added. Specifically, each decoding layer in the network expands a branch composed by a deconvolution and a $1\times1$ convolution to up-sample the feature maps to the same resolution as input and squeeze the output channels. Then the training labels are added for the supervised learning (see the thinner orange arrows in Fig.~\ref{fig1}). This allows an introduction of additional gradients during training and further alleviates the vanishing of gradients. To some extent, it can be also regarded as an additional constraint for the network to avoid overfitting. Finally, the between-net connections are designed. The features from the decoding layers of the first U-Net are transmitted to the corresponding encoding layers in the second U-Net by concatenation operation. These between-net connections enable the high-resolution information in shallow layers to be preserved and sent to the deeper layers for a fine segmentation of tumor substructures.

\subsection{Training with Loss Weighted Sampling}
Our proposed network is an end-to-end architecture, in which the two cascaded U-Nets are trained jointly, ensuring the efficiency of the data processing procedure. To address the extremely imbalance of the positive and negative samples in brain tumor dataset, we present a loss weighted sampling scheme and introduce it into the cross entropy loss function. Specifically, the sampled loss is formulated as follows:
\begin{equation}
\mathcal{L}=\frac{\sum\limits_{n=1}^{b}\sum\limits_{i=1}^{l}\sum\limits_{j=1}^{w}\left[\left(-\sum\limits_{m=1}^{c}\left(L \cdot \log{Y} \right)\right)\cdot W\right]}{\sum\limits_{n=1}^{b}\sum\limits_{i=1}^{l}\sum\limits_{j=1}^{w} W}\label{1}
\end{equation}
where $Y\!\in\!\mathbb{R}^{b\times c\times l\times w}$ denotes the predicted probability for the one-hot label $L\in\mathbb{R}^{b\times c\times l\times w}$ after softmax functions. $b$ is the number of batches, $c$ is the number of channels, $l$, and $w$ are the length and width of the image, respectively. Sample matrix $W \!\in\!\mathbb{R}^{b\times l\times w}$ is computed according to specific tasks, and $W_{n,i,j} \in \{0,1\}$ denotes the loss weight of the pixels at the spatial location $(n,i,j)$.

\begin{figure*}[htpb]
\setlength{\belowcaptionskip}{-0.1cm}
\centering
\includegraphics[width=12.6cm, height=3.9cm]{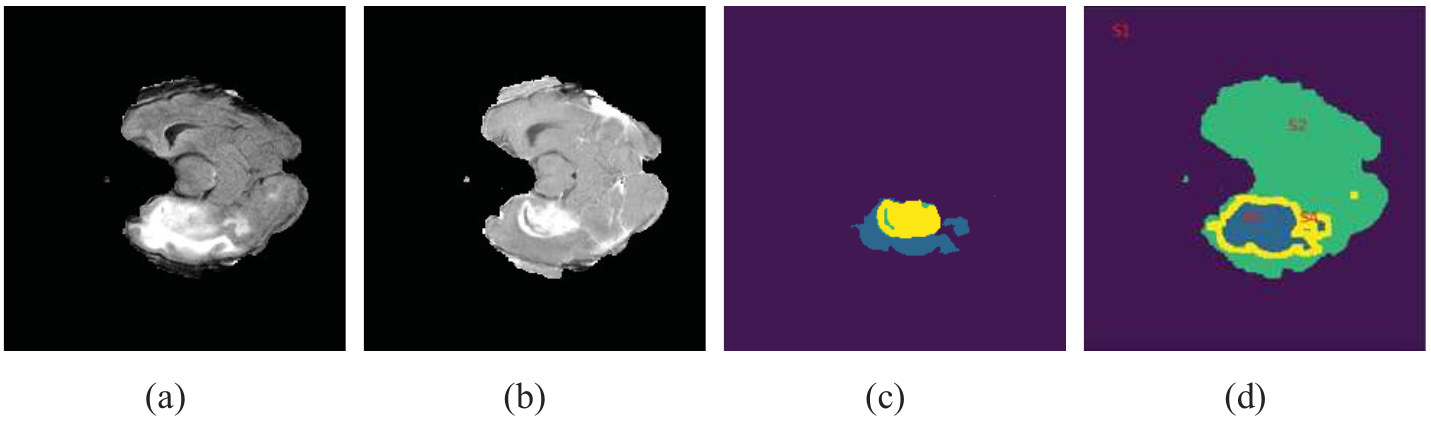}
\vspace{2mm}
\caption{A brain tumor training sample is divided into four regions according to the input data and ground truth: (a) FLAIR. (b) T1ce. (c) Ground truth (Purple: Non-tumor; Blue: Edema; Yellow: enhancing tumor; Green: necrosis.) (d) Four regions of a training sample. $S_{1}$: Black background; $S_{2}$: Normal brain region; $S_{3}$: Tumor region obtained from (c); $S_{4}$: Tumor contour region obtained by a contour detection algorithm.}
 \label{fig2}
\end{figure*}

The brain MRI image is divided into four regions: $S_{1}, S_{2}, S_{3}$ and $S_{4}$, which represent the black background, normal brain region, tumor region, and tumor contour region, respectively (see Fig.~\ref{fig2} (d)). Then the sample matrix $W$ can be computed as:
\begin{equation}
W = \sum\limits_{i=1}^{3}Sample\left(S_{i},p_{i}\right)+\alpha Sample\left(S_{4},p_{4}\right)\label{2}
\end{equation}
where $Sample(S_{i},p_{i})$ denotes a binary matrix obtained by random sampling in $S_{i}$ with probability $p_{i}$. The hyper-parameter $\alpha$ is greater than or equal to 1, which is introduced for adjusting the loss weight of contour regions and is expected to enhance the ability of network to recognize the tumor contour.

For most of the MRI images, the black background $S_{1}$, also referred as non-brain mask in this paper, contains a large number of pixels, but provides little useful information for segmentation of the tumor. So according to this prior knowledge, we let $p_{1}$ be 0 and extract a non-brain mask in advance and merge it with the prediction maps when testing.

To compute the branch loss $\mathcal{L}_{1}$ and auxiliary loss $\mathcal{L}_{a_i}\!$ $(i=1,2,\ldots,4)$ in U-Net1, we let $p_{3} = 1, p_{4} = 1$. Then $p_2$ is calculated by:
\begin{equation}
p_{2}\cdot N_{S_{2}} = \beta \cdot p_{3}\cdot N_{S_{3}}\label{eq3}
\end{equation}
where $N_{S_i}$ denotes the pixel number in region $S_{i}$, and $\beta$, usually more than 1, is for adjusting the proportion of positive and negative samples in a training batch, thus eliminating the class imbalance. Because $Sample(S_i,p_i)$ is a random sampling operation, as long as $\beta\cdot p_{2}\cdot epoch \geq 1$ is guaranteed, where $epoch$ is the times of the network to pass whole training set, all pixels in the dataset are expected to participate in the calculation of loss for at least one time so that no information from the brain tumor will be lost.

For the branch loss $\mathcal{L}_{2}$ and auxiliary loss $\mathcal{L}_{a_i}\!$ $(i\!=\!5,6,\ldots,8)$ in U-Net2, we let: $p_{1}\!=\!0, p_{2}\!=\!0, p_{3}\!=\!1, p_{4}\!=\!1, \alpha_{2}\!=\!1$, which means that U-Net2 only learns the segmentation of tumor substructures. Thus, the loss function of our network is
\begin{equation}
\mathcal{L}_{Total}= \mathcal{L}_{1}+ \mathcal{L}_{2} + \omega\sum\limits_{i=1}^{8}{\mathcal{L}_{ai}}+\lambda \psi\label{4}
\end{equation}
where $\mathcal{L}_{a}$ is the auxiliary loss, $\omega$ is the weighted coefficient, and $\psi$ is the regularization term with hyper-parameter $\lambda$ for tradeoff with the other terms.

For the testing process, we extract the non-brain mask in advance and fuse it with the outputs of branch1 and branch2 to get the final segmentation result.

\section{Experimental Results}
\subsection{Datasets and Pre-processing}
We evaluate our method on the training data of BraTS challenge 2017. It consists of 210 cases of high-grade glioma and 75 cases of low-grade glioma. In each case, four modal brain MRI scans: T1, T2, T1ce and FLAIR, are provided, respectively. The resolution of MRI scans is $240\times240\times155$. Pixel-level labels provided by the radiologists are: 1 for necrotic (NCR) and the non-enhancing tumor (NET), 2 for edema (ED), 4 for enhancing tumor (ET), and 0 for everything else. In our experiments, 210 high-grade cases  are divided into three subsets at a ratio of $3:1:1$, i.e., 126 training data, 42 validation data and 42 testing data are attained. Low-grade cases are not used. Besides, about 30\% scans that don't contain any tumor structure are discarded in the training process. All the input images are processed by N4-ITK bias field correction and intensity normalization. Data augmentation including random rotation and random flip is used in all algorithms.

\subsection{Implementation Details}
All the algorithms were implemented on a computer with NVIDIA GeForce GTX1060Ti (6 GB) GPU and Intel Core i5-7300HQ CPU @ 2.5 GHz (8GB), together with the open-source deep learning framework pytorch. The contour weight $\alpha_{1}$ is set to 2, and $\beta$ is set to 1.5. The extracted tumor contour is about 10 pixels wide. In the training phase, we use
stochastic gradient descent (SGD) with momentum to optimize the loss function. The momentum parameter is 0.9, learning rate of $10^{-3}$ initially and decreased by a factor of 10 every ten epochs until a minimum threshold of $10^{-7}$. The weight decay $\lambda$ is set to $5\times10^{-5}$. The models are trained for about 50 iterations until there is an obvious uptrend in the validation loss. The weighted coefficient $\omega_{a}$ is set to 0.1 initially and decreased by a factor of 10 every ten epochs until a minimum threshold of $10^{-3}$.

For segmentation results, we evaluate the following three parts: (1) Whole Tumor (WT); (2) Tumor Core (TC); and (3) Enhancing Tumor (ET). For each part, Dice score, sensitivity and specificity are defined as follows:
\begin{equation}
Dice(P,T)=\frac{2|P_{1}\wedge T_{1}|}{|P_{1}|+|T_{1}|}, Sens(P,T)=\frac{|P_{1}\wedge T_{1}|}{|T_{1}|}, Spec=\frac{|P_{0}\wedge T_{0}|}{|T_{0}|}\label{5}
\end{equation}
where $P$, $T$ denote the segmentation results and labels, and $P_{0}, P_{1}, T_{0}, T_{1}$ denote negatives in $P$, positives in $P$, negatives in $T$ and positives in $T$, respectively.

\subsection{Results and Analysis}
To verify the effectiveness of our proposed network and loss weighted sampling scheme, we compare our CUN method with several state-of-the-art deep learning algorithms including U-Net~\cite{ronneberger2015u}, BFCN~\cite{shen2017boundary} and DRN~\cite{lopez2017dilated}.

\begin{figure*}[htbp]
\setlength{\belowcaptionskip}{-0.1cm}
\centering
\includegraphics[width=0.986\columnwidth]{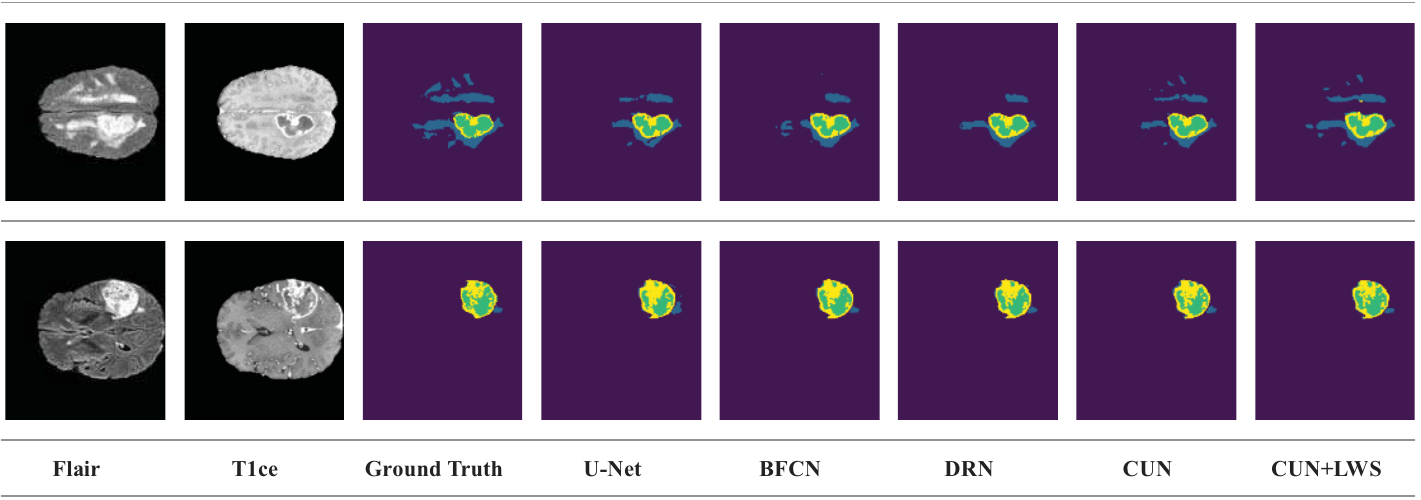}
\vspace{2mm}
\caption{Segmentation results of different methods on the local testing data. From left to right: Flair, T1ce, Ground Truth and results of U-Net, BFCN, DRN, CUN, CUN+LWS, respectively. In ground truth and segmentation results, purple, blue, yellow, green represent Non-tumor, Edema, Enhancing Tumor, and Necrosis, respectively.}
\label{fig3}
\end{figure*}

\begin{figure*}[htpb]
\setlength{\belowcaptionskip}{-0.1cm}
\centering
\includegraphics[width=12.6cm, height=6.6cm]{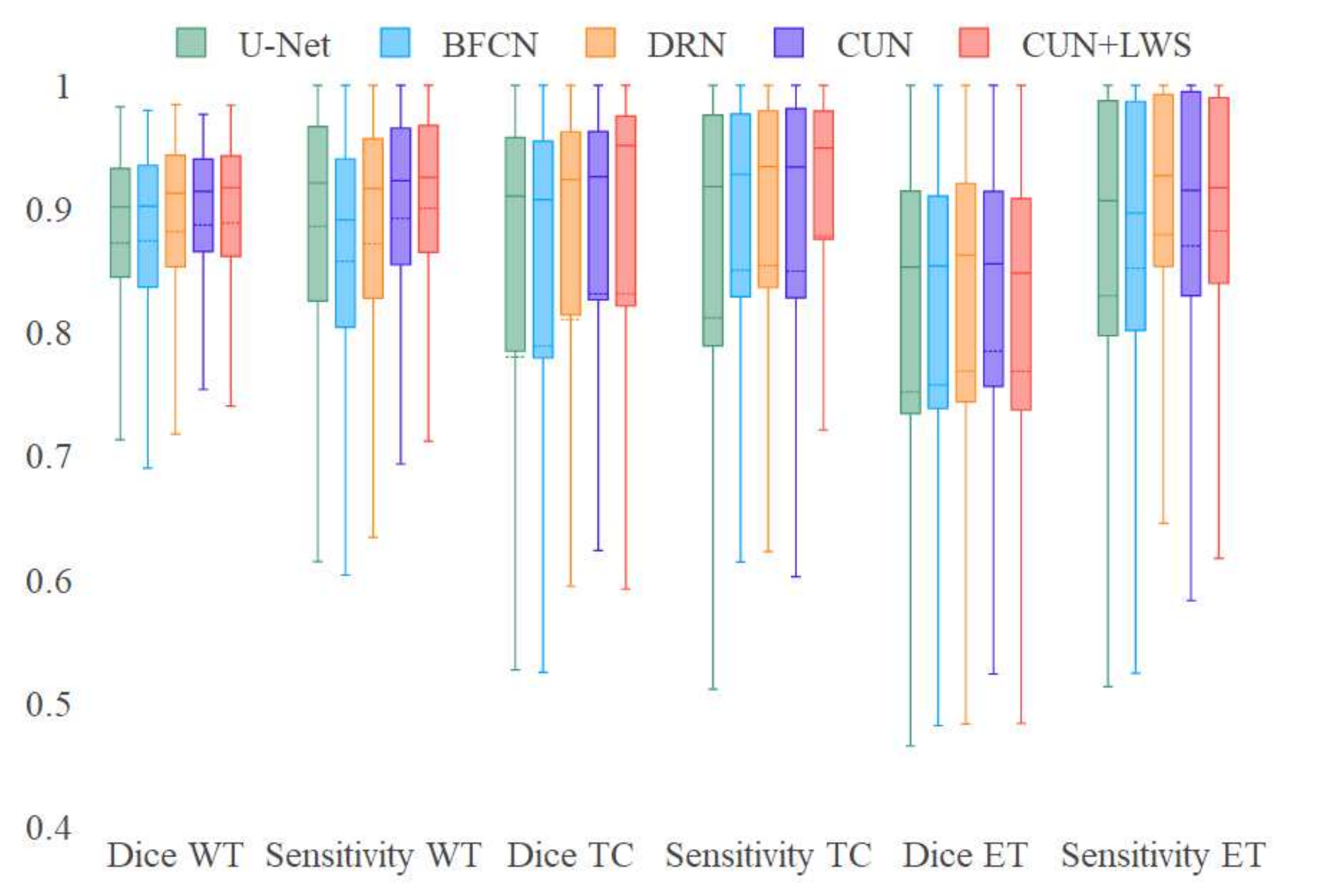}
\vspace{2mm}
\caption{Distributions of dice score and sensitivity of the five methods for whole tumor (WT), tumor core (TC) and enhancing tumor (ET). The solid lines and dotted lines in the boxes represent the median value and the average value, respectively.}
\label{fig4}
\end{figure*}

The visual results are shown in Fig.~\ref{fig3}. It can be seen that our proposed CUN+LWS has the best segmentation sensitivity among the five methods, and is better at segmenting the tiny sub-structures within a brain tumor. The distributions of the obtained Dice scores and sensitivities are presented in Fig.~\ref{fig4}. The quantitative results of the five models on the testing set are listed in Table~\ref{tab1}. As we can see, our CUN method outperforms the three state-of-the-art methods by approximately 1.5 percent in dice score and 2 percent in sensitivity. Besides, when LWS is adopted, there is an additional average growth of 1.5 percent in sensitivity, which indicates the effectiveness of LWS.

\begin{table*}[htpb]
\normalsize
\setlength{\belowcaptionskip}{-0.3cm}
\centering
\caption{Dice score, sensitivity and specificity of the five methods for Whole Tumor, Tumor Core and Enhancing Tumor over the testing set.}
\label{tab1}
\renewcommand\arraystretch{1.25}
\resizebox{\textwidth}{17.5mm}{
\begin{tabular}{@{}lcccccccccc@{}}
\toprule
\multirow{2}{*}{Method}& \multicolumn{3}{c}{Whole Tumor}                  & \multicolumn{3}{c}{Tumor Core}       & \multicolumn{3}{c}{Enhance Tumor}\\ \cmidrule(l){2-10}
                       & \textbf{Dice}  & \textbf{Sens}   & \textbf{Sper} & \textbf{Dice}   & \textbf{Sens}& \textbf{Sper}& \textbf{Dice}   & \textbf{Sens} & \textbf{Sper}  \\
\midrule
U-Net~\cite{ronneberger2015u}                  & 0.872          & 0.885          & 0.996         & 0.779         & 0.811        & \textbf{0.999}& 0.751         & 0.829      & \textbf{0.999}\\
BFCN~\cite{shen2017boundary}     & 0.874         & 0.857         & \textbf{0.997}& 0.788         & 0.850          & 0.998       & 0.757         & 0.851          & 0.998\\
DRN~\cite{lopez2017dilated}                    & 0.881         & 0.871       & 0.996  & 0.810         & 0.854       & 0.998     & 0.768         & 0.878          & 0.998\\
CUN                  & 0.886        & 0.892        & \textbf{0.997}  & 0.830         & 0.849    & \textbf{0.999}    & \textbf{0.784}& 0.869          & 0.998\\
CUN+LWS                & \textbf{0.888} & \textbf{0.903}& 0.996  & \textbf{0.831} & \textbf{0.877} & 0.998        & 0.768        & \textbf{0.881}& 0.998\\
\bottomrule
\end{tabular}}
\end{table*}

\section{Conclusions}
Inspired by the hierarchical structure of brain tumors, we proposed a novel cascaded U-Net for the segmentation of brain tumor. To make the network work more effectively, three strategies were designed. The residual blocks and the auxiliary supervision can help gradient flow more smoothly during training, and alleviate the gradient vanishing problem caused by increasing network depth. The between-net connections can transmit the high resolution information from the shallow layer to the deeper layer and obtain more refined segmentation results. Furthermore, we presented a loss weighted sampling scheme to adjust the number of samples in different classes to solve the severe class imbalance problem. Our experimental results demonstrated the advantages of our network and the effectiveness of the loss weighted sampling scheme.

\section*{Acknowledgments}
This work was supported by the Project supported the Foundation for Innovative Research Groups of the National Natural Science Foundation of China (No.\ 61621005), the Major Research Plan of the National Natural Science Foundation of China (Nos.\ 91438201 and 91438103), the National Natural Science Foundation of China (Nos.\ 61876221, 61876220, 61836009, U1701267, 61871310, 61573267, 61502369 and 61473215), the Program for Cheung Kong Scholars and Innovative Research Team in University (No.\ IRT\_15R53), the Fund for Foreign Scholars in University Research and Teaching Programs (the 111 Project) (No.\ B07048), and the Science Foundation of Xidian University (Nos.\ 10251180018 and 10251180019).

\bibliographystyle{splncs04}
\bibliography{references}

\end{document}